# Analysis of incubation time preceding the Ga-assisted nucleation and growth of GaAs nanowires on Si(111)

Faebian Bastiman[1], Hanno Küpers[1], Claudio Somaschini[1,†], Vladimir G. Dubrovskii[2], and Lutz Geelhaar[1,*]

[1]*Paul-Drude-Institut für Festkörperelektronik, Leibniz-Institut im Forschungsverbund Berlin e.V., Hausvogteiplatz 5-7, 10117 Berlin, Germany*

[2]*ITMO University, Kronverkskiy pr. 49, 197101 St. Petersburg, Russia*

[†]*Present address: PoliFab – Politecnico di Milano, via Giuseppe Colombo 81, 20133 Milan, Italy*

\* geelhaar@pdi-berlin.de

ABSTRACT: The incubation time preceding nucleation and growth of surface nanostructures is interesting from a fundamental viewpoint but also of practical relevance as it determines statistical properties of nanostructure ensembles such as size homogeneity. Using *in situ* reflection high-energy electron diffraction, we accurately deduce the incubation times for Ga-assisted GaAs nanowires grown on unpatterned Si(111) substrates by molecular beam epitaxy under different conditions. We develop a nucleation model that explains and fits very well the data. We find that, for a given temperature and Ga flux, the incubation time always increases with decreasing As flux and becomes infinite at a certain minimum flux, which is larger for higher temperature. For given As and Ga fluxes, the incubation time always increases with temperature and rapidly tends to infinity above 640 °C under typical conditions. The strong temperature dependence of the incubation time is reflected in a similar variation of the nanowire number density with temperature. Our analysis provides understanding and guidance for choosing appropriate growth conditions that avoid unnecessary material consumption, long nucleation delays, and highly inhomogeneous ensembles of nanowires. On a more general ground, the existence of a minimum flux and maximum temperature for growing surface nanostructures should be a general phenomenon pertaining for a wide range of material-substrate combinations.

## I. INTRODUCTION

Semiconductor nanowires (NWs) are believed to be essential building blocks for future electronic and optoelectronic applications [1]. One specific advantage is that epitaxy in the form of NWs enables monolithic integration of III-V compound semiconductors on Si substrates in high crystalline quality, because strain induced by lattice mismatch can elastically relax at the free sidewall surfaces [2] and dislocations terminate there [3]. This particular benefit may lead to efficient light emitters for the Si platform, which could be the basis for future on-chip optical data communication [4], and advanced solar cells [5]. While there are several approaches for the bottom-up growth of III-V compound semiconductor NWs, vapor-liquid-solid growth from droplets of the group-III metal has become very popular, since it avoids the use of any foreign materials that could be detrimental for device performance [6]. Among different variations of this approach, the Ga-assisted growth of GaAs NWs by molecular beam epitaxy (MBE) has been investigated most intensively [7].

Understanding and controlling the formation of NWs is a mandatory prerequisite for exploiting their benefits in applications. A particularly crucial aspect in this context is NW nucleation because this stage of the entire formation typically determines the position and diameter of the NWs. For Ga-assisted GaAs NWs, nucleation is a fairly complex process that involves the formation of Ga droplets, the infusion and saturation of these droplets by incident As, the nucleation of an initial GaAs seed crystal under the droplet, and the evolution of the seed into a NW with the droplet seated on the top facet. These are the steps that are essential for an idealized NW nucleation event, but in reality other processes are also relevant. First, for the Ga-assisted growth of GaAs NWs Si substrates are typically employed that are covered with a thin $SiO_x$ layer to facilitate the formation of Ga droplets. Droplet nucleation is enhanced in pin holes of the oxide layer [8–10]; and in addition the Ga droplet may react with both the oxide layer and Si substrate such that the oxide is locally dissolved [11] and the interface with the substrate roughens [12] or GaAs even grows into the Si substrate [13]. Second, multiple twinning of the initial GaAs seed under the droplet may lead to the formation of inclined and horizontal NWs [13,14]. Third, rapid consumption of the Ga droplet may result in the formation of a GaAs crystallite instead of a GaAs NW [15]. Alternatively, GaAs may not form at all under the droplet if locally the As flux is too low [15]. All objects that are not vertical NWs are typically considered parasitic. The yield of vertical NWs is largely determined during the nucleation stage, which contributes to its paramount importance.

Naturally, any of the above individual steps may be affected by the choice of growth conditions. At the same time, the elucidation of the detailed interdependencies presents a formidable challenge, and ideally one would like to follow such processes *in situ*. A parameter that can be probed *in situ* fairly easily is the incubation time. In general, any nucleation may be preceded by an incubation time that elapses after the supply of precursor species is started. This phenomenon has been studied extensively for the self-assembled growth of GaN NWs (without any droplets), where the incubation step can last more than one hour [16–22]. An incubation time before NW nucleation has also been reported for the vapor-liquid-solid growth of Si and Ge as well as the vapor-solid-solid growth of GaN and ZnTe NWs, and was attributed to the time needed to reach the critical supersaturation in the droplets [23–28] and particles [29–31], respectively.

For the Ga-assisted formation of GaAs NWs, already in one of the first publications on this topic the extrapolation of a NW length series to zero implied that nucleation is preceded by an incubation time also in this case [8]. However, it was only more recently that this incubation time was directly monitored *in situ* by reflection high-energy electron diffraction (RHEED) [32,33]. It was found that the incubation time depends on the thickness of the oxide layer [33], which underlines the important influence of the latter on the nucleation of Ga-assisted GaAs NWs. A theoretical analysis of the initial Ga droplet size was based on the assumption that an incubation time is needed to reach a critical As concentration in the droplets [34]. Further studies showed how the incubation time affects the length distribution of the final NW ensembles [35,36]. Faster nucleation always yields more uniform length over the ensemble, while longer nucleation delay results in broader length distributions with a pronounced asymmetry toward smaller lengths (corresponding to nanowires that have emerged later). However, full understanding of the incubation time preceding the Ga-assisted growth of GaAs NWs is still lacking.

Consequently, here we report a comprehensive experimental and theoretical investigation of how the incubation time of Ga-assisted GaAs NWs depends on the growth conditions such as the growth temperature and material fluxes. We accurately deduce the incubation times under different conditions and find the following general trends. The incubation time for GaAs NWs equals the sum of two constituent times, where the first is required to nucleate Ga droplets and the second to start the NWs themselves by the precipitation of GaAs. For a given temperature and Ga flux, the incubation time always increases with decreasing As flux and becomes infinite at a certain minimum flux, which is larger for higher temperature. For a given As and Ga flux, the incubation time always increases with increasing temperature and becomes infinite above 640 °C under typical conditions. The temperature dependence of the incubation time is extremely steep. It can be approximated by the Arrhenius-type function for temperatures lower than 620 °C, while for higher temperatures it increases much more rapidly. This peculiar behavior is reflected in the temperature dependence of the total density of all objects nucleated from Ga droplets after the end of growth. A dedicated model explains and fits very well the measured incubation times versus temperature and As flux.

## II. EXPERIMENTAL METHODS

GaAs NWs were grown in the Ga-assisted mode by solid-source MBE on 2" n-type Si(111) substrates. Prior to growth, substrates were annealed in the growth chamber at (695±5) °C for 20 min, and then the substrate temperature was decreased to the desired growth temperature $T$, which was measured by a pyrometer. Next, the substrate was exposed for 5 min to As$_4$ before growth was started by opening the Ga shutter. In a systematic series of experiments, the $As_4$ flux was varied in the range 0.69±0.03 to 18.5±0.9 atoms/nm$^2$/s and $T$ in the range 550 to 640 °C, while the Ga flux and the growth duration were always fixed at 0.69±0.03 atoms/nm$^2$/s and 30 min, respectively. A few growth runs were repeated under identical conditions to verify reproducibility. These experiments were carried out in the same way as the more comprehensive variation of growth conditions described in our previous publication [15], which contains more information on calibration procedures and further experimental details. For comparison, we also made use of growth experiments in which Ga was pre-deposited before $As_2$ was supplied so that Ga droplets formed before GaAs could nucleate (We did not observe any differences between experiments with $As_2$ and $As_4$). These experiments are described in more detail in another previous publication [37].

In order to investigate the nucleation process *in situ*, RHEED videos were recorded during the initial phase of growth with substrate rotation and analyzed later with the software Safire. The morphology of the resulting samples was characterized by scanning electron microscopy (SEM). Since the morphology of all samples varied systematically from the wafer center to the edge, only the same central region was always measured. Images were analyzed with the software ImageJ to determine in a reliable way the number density of all objects on the surface. Different types of objects were distinguished on the basis of top-view and 10° off top-view images, and in some cases individual number densities were extracted by comparing the overgrown surface area with the average cross-section of individual objects [15]. The NW elongation rate was deduced by dividing the average NW length obtained from cross-sectional micrographs by the growth duration.

## III. EXPERIMENTAL RESULTS

The central experimental idea of this study is to measure the incubation time of Ga-assisted GaAs NWs by RHEED as a function of growth conditions. The evolution of the RHEED pattern monitored during nucleation and growth is illustrated by the four images in the inset of Fig. 1. When the sample is annealed under an As overpressure before any Ga is supplied, a diffuse intensity distribution is recorded, and the only discernible feature is a faint specular reflection (image i). Such a pattern is characteristic for scattering from an amorphous layer, such as the native Si oxide that covers the Si substrate. At some time after exposure of the substrate to the Ga flux, an additional reflection appears on the zeroth-order rod (image ii). In some cases, this reflection can be difficult to discern. The moment in time at which this pattern occurs depends on the growth conditions. Within 2–4 further seconds, the RHEED pattern characteristic for electron transmission through GaAs NWs [38,39] forms (image iii). The arrangement of reflections reveals that GaAs nucleates in both wurtzite and zincblende crystal phases, with the latter being present in the two twinned orientations. As the GaAs growth proceeds, this pattern increases in intensity, and depending on growth conditions the relative intensity of the reflections corresponding to the two crystal phases may change. Image iv depicts an example of a RHEED pattern acquired after the growth of well-developed NWs. In this case, the zincblende phase is clearly dominant, as it is typical for most Ga-assisted GaAs NWs.

For all further analyses, the time between the opening of the Ga shutter and the appearance of pattern iii is chosen as the incubation time. In principle, the analysis could have been based also on the occurrence of pattern ii, which may be related to the nucleation of the first GaAs monolayer(s) under the Ga droplets. However, pattern iii develops in much higher contrast and is hence easier to discern. Also, the interval between the two events was for all samples considered in this work fairly constant (2–4 s). The RHEED pattern iii appears six times per substrate revolution, which corresponds to once per second at the substrate rotation speed employed in our experiments. Hence, the incubation times have an inherent error of ±1 s. However, in the case of samples where the total number density of all crystalline objects, i.e. vertical NWs, inclined NWs as well as crystallites, is very low, the occurrence of the RHEED pattern is initially less distinct. Thus, a relative error of ±10% is assumed for the incubation time. Furthermore, this analysis is sensitive to the nucleation of GaAs in general and not specific to NWs. This aspect could be important since in addition to NWs, parasitic growth of GaAs on the substrate may be substantial, depending on the growth conditions. However, it has been shown that all GaAs objects originally nucleate from Ga droplets [15,40], and there is no reason to expect a systematic delay between the nucleation of different objects. Hence, the analysis approach chosen here is expected to provide meaningful data. As a last note of caution, we mention that the Ga-assisted formation of GaAs NWs is known to depend sensitively on properties of the Si oxide [10], and thus the substrate preparation may affect all quantitative values reported in the following.

Fig. 1(a) presents the variation of the incubation time with As flux for $T$=580 °C and 620 °C. With increasing As flux, the incubation time decreases, and it increases with temperature. Generally, the incubation time before nucleation of GaAs NWs is the sum of the nucleation time for Ga droplets on the substrate and the nucleation time for GaAs seeds below these droplets. We note that in this consideration, any time needed for dissolution of Si oxide is included in the droplet formation, since these two processes take place simultaneously. The observed dependence on the As flux is expected for the nucleation time of GaAs: with increasing As flux the supersaturation inside the Ga droplets rises, and hence the probability for nucleation of GaAs. Thus, the incubation time is shortened. The variation of the incubation time with $T$ will be described in more detail below. Furthermore, the data for $T$=620 °C exhibit a drastic rise in incubation time below an As flux of 2.8 atoms/nm$^2$/s (please note the logarithmic scale), and a similar behavior is evident for 580 °C at 1.7 atoms/nm$^2$/s. For $T$=620 °C, the incubation time varies below the critical As flux of 2.8 atoms/nm$^2$/s by over an order of magnitude. The divergence of the incubation time toward larger values for lower As fluxes is consistent with the earlier observation that a minimum As flux is needed for NW elongation [15,41,42], and that this flux increases with $T$ [15]. This effect was attributed to As desorption from the droplet. Apparently, the nucleation probability decreases significantly as the As flux approaches the minimum required flux. Above the critical As flux, the dependence of the incubation time on As flux can approximately be described for both $T$ by an exponential function, as indicated by the dashed lines. Further below in the modelling section, we will develop a proper functional description of the experimental data in the entire range of As fluxes. However, the exponential approximation enables the assignment of a critical value below which the variation in incubation time becomes much more rapid.

Independent of the detailed mathematical description, the change of the incubation time with As flux implies a variation in the Ga droplet size at the point of nucleation, as explained in the following. From our previous study [15] we know that the number density of all objects on the sample does not depend on As flux. In addition, we deduced that all objects form from the initial Ga droplets. Hence, the number density of Ga droplets is constant for a given T. If now the incubation time is increased due to a change in As flux, more Ga accumulates in the same number of Ga droplets, and consequently they must increase in size.

In order to deduce the dependence of the incubation time on *T*, samples grown with too low As flux must clearly be avoided. For *T*=620 °C, the lowest As flux that can be used is 2.8 atoms/nm$^2$/s, i.e. the critical value. Since for lower *T* the critical As flux is lower, it is safe to focus on the samples grown with this As flux. The resulting dependence of the incubation time on *T* is presented in Fig. 1(b). The incubation time increases with *T*, which is in agreement with the difference between the data sets for the two *T* in Fig. 1(a). Beyond this simple monotonous dependence, the data in Fig. 1(b) exhibits a rapid rise at a critical temperature of 620 °C. This aspect is more easily seen on the Arrhenius plot in the inset. This plot shows that the variation of the incubation time on the lower temperature side of the critical temperature can approximately be fitted by the Arrhenius dependence with an activation energy of (1.2±0.1) eV. Above the critical temperature the variation is much stronger. Therefore, for both the As flux and temperature dependences of the incubation time, the nucleation processes cannot be described by simple exponential functions over the entire range. Rather, beyond a certain critical As flux or temperature, the incubation time starts to increase very abruptly, implying that practically no growth can occur outside a certain range of these growth parameters. Again, a complete description of the data will be developed in the modelling section.

We note that in our previous study, we deduced independently the same critical temperature of 620 °C on the basis of the sample morphology [15]. In particular, we found that above 620 °C the inhomogeneity of the length and diameter within the NW ensembles drastically increases. The broadening of the length and diameter distributions can qualitatively be explained by differences in the nucleation process as follows. A large variation of the NW diameter implies that the diameters of the original Ga droplets exhibit a broad distribution. Also, the simplest explanation for a pronounced inhomogeneity of the NW length is that NWs nucleate at different times [35,36], leading to broader length distributions for longer incubation times. Conversely, NWs that are very homogeneous in length and diameter, as found below 620 °C, must have nucleated from similar droplets and almost simultaneously. We can thus conclude that for lower temperatures (at a given As flux inside the optimum range as discussed above), Ga droplets form quickly on the substrate and the NW nucleation is synchronized in time.

In contrast, above 620 °C the inhomogeneity of the NW length distribution is on the same order as the average NW length. Hence, the incubation time must vary on the same order as the total growth duration, which is much longer than the incubation time indicated by the occurrence of the first GaAs reflection in RHEED as displayed in Fig. 1(b). Therefore, in this regime the long overall incubation time must be related not to a delay in GaAs precipitation below the Ga droplets but rather to the nucleation of the droplets themselves, i.e. to the other component of the total incubation time for GaAs NWs. At these high temperatures, Ga sticking is significantly reduced [43], which changes the balance of atomic processes occurring on the surface. This conclusion is in agreement with our previous observation that above 640 °C no growth takes place at all [15]. Furthermore, the indication of nucleation in the RHEED patterns is less distinct for the highest $T$, which is consistent with the conclusion that in these cases the nucleation of GaAs under different droplets occurs over an extended time. Therefore, we attribute the rapid increase of the incubation time, observed above 620 °C, to the nucleation process becoming limited by the Ga desorption. This limitation prevents that the nucleation of the entire ensemble of Ga droplets is completed before the first precipitation of GaAs.

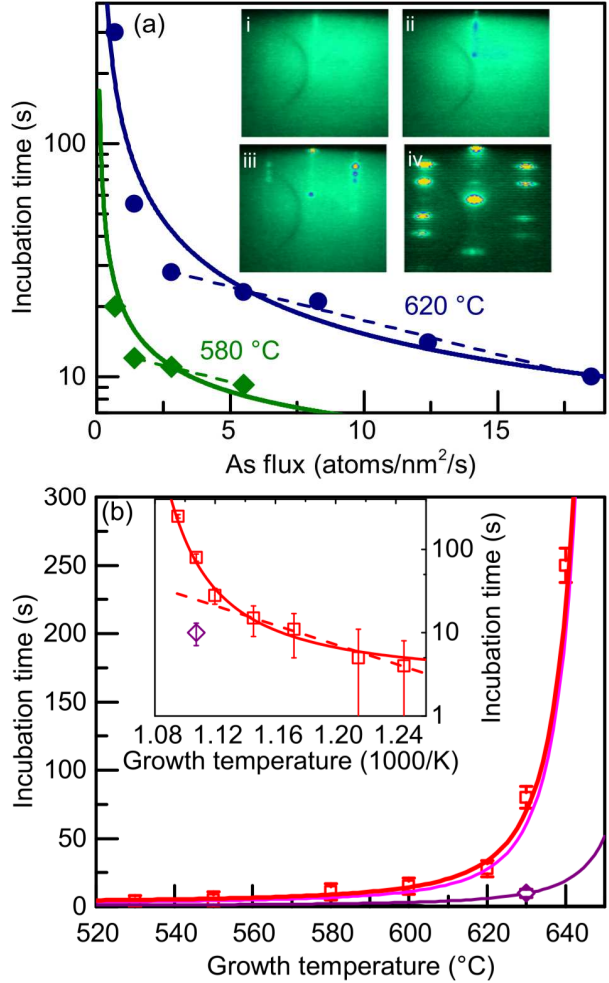

FIG. 1. (a) Dependence of the incubation time on As flux at $T$=580 °C (diamonds) and 620 °C (circles). The dashed lines indicate the exponential fits. The inset depicts the evolution of the RHEED pattern during nucleation and growth of Ga-assisted GaAs NWs on Si(111) in different stages: (i) Prior to supplying Ga. (ii) At some point after opening the Ga shutter, an additional reflection appears on the zeroth-order rod. The elapsed time depends on the growth conditions. (iii) 2–4 s later, the pattern characteristic for GaAs NWs forms. (iv) Typical pattern after the growth of well-developed NWs. (b) Dependence of the incubation time on $T$ at an As flux of 2.8 atoms/nm²/s. The insert shows the same data in an Arrhenius plot, revealing an activation energy of (1.2±0.1) eV to the right of the critical temperature (corresponding to the dashed line). The purple data point (diamond) indicates the average incubation time for experiments in which Ga was pre-deposited prior to supplying As and the incubation time was counted starting from supplying As, such that the formation time of Ga droplets was not included in the measurement. In both panels, the solid lines show theoretical fits based on the model presented in the main text, describing the non-linear behavior of the incubation time versus the As flux and temperature over the entire range of data. The purple line in (b) corresponds to the incubation time of Ga droplets and the magenta line to the incubation time of GaAs NWs.

If our conclusion is true that the formation of Ga droplets is different below and above 620 °C, this difference should also be apparent in the variation of the droplet number density with $T$. Instead of measuring the droplet number density on dedicated samples without NWs, we make use of the fact that this density corresponds to the total density of all objects present on the substrate surface after NW growth [15,40]. In a previous study, we also observed that the latter density does not depend on As flux [15]. For the present investigation, we focus again for each $T$ on the samples with an As flux of 2.8 atoms/nm$^2$/s. The resulting dependence of the total number density of all objects on $T$ is shown in Fig. 2. As expected, this curve exhibits a substantial change in slope at the same critical temperature of 620 °C as found for the incubation time in Fig. 1(b). Again, the change in the dependence at the critical temperature is more clearly seen on the Arrhenius plot in the inset to Fig. 2. The dependence is exponential on the lower temperature side of the critical temperature, with an activation energy of (1.2±0.1) eV. At higher temperatures the variation is much stronger, similarly to Fig. 1(b).

Very remarkably, the shapes of the curves in Figs. 2 and 1(b) are very similar and the critical temperatures are identical. This agreement implies that the incubation time for Ga-assisted GaAs NWs is governed by the Ga droplet formation process. In other words, as long as the As flux is sufficiently high, the precipitation of GaAs below the Ga droplets is fast compared to the nucleation of the droplets themselves. Only the small variation of the incubation time with As flux at a given *T* seen in Fig. 1(a) is related to the time required to nucleate GaAs after droplet formation, and as mentioned above the dependence is as expected.

In order to verify this conclusion, we make use of experiments in which Ga was pre-deposited prior to providing As such that droplet formation and GaAs precipitation are decoupled [37]. These experiments were carried out at *T*=630 °C. The incubation time was determined in the same way as described above, but starting from opening the As rather than the Ga shutter. We did not observe any significant difference in incubation time between experiments with 30 and 60 s of Ga pre-deposition time. The resulting average value of 10 s is also plotted in Fig. 1(b). Clearly, this incubation time is much shorter than for the experiment without Ga pre-deposition, which confirms that in the latter case the incubation time is limited by the formation of droplets.

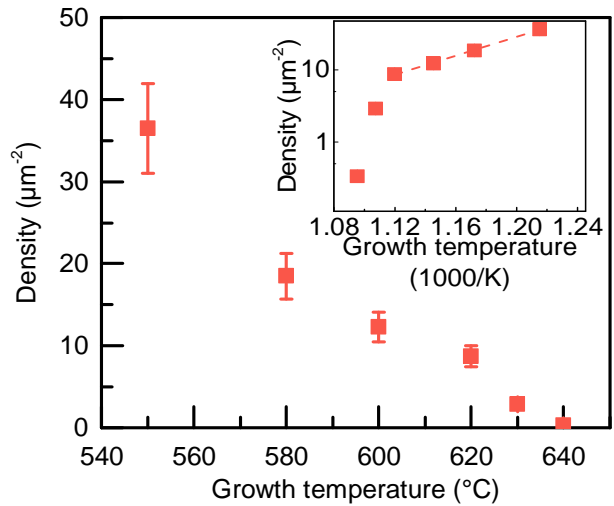

FIG. 2. Total density of all objects versus *T* at an As flux of 2.8 atoms/nm²/s. The inset shows the same data in an Arrhenius plot, revealing an activation energy of (1.2±0.1) eV to the right of the critical temperature (dashed line).

## IV. MODELLING

We now present a model to quantitatively describe the incubation times versus the As flux and temperature over the entire range of the obtained data. From Figs. 1(a) and (b), the incubation time increases with decreasing As flux and increasing temperature, and this increase becomes much more rapid when these parameters reach some critical values. According to the above discussion, the incubation time in the general case contains two contributions, the nucleation time for Ga droplets and for GaAs NWs themselves, meaning that the total incubation time can be limited by one of the two processes under different conditions. As discussed above, the As flux dependences shown in Fig. 1(a) are governed by the nucleation time of GaAs NWs, while the temperature dependence in Fig. 1(b) is mainly driven by the nucleation time for Ga droplets.

According to nucleation theory, any material $k$ ($k$ = Ga and GaAs, where nucleation of GaAs NWs from Ga droplets is limited by the As flux) can start growing only if its concentration $c_k$ is larger than the equilibrium, $c_k > c_k^{eq}$ [44]. The temperature dependence of the equilibrium concentration is Arrhenius-like in the first approximation, i. e., $c_k^{eq} = c_k^0 \exp(-\Lambda_k / k_B T)$, with $\Lambda_k$ as the specific nucleation heat, $k_B$ as the Boltzmann constant, and a temperature-independent $c_k^0$ [44]. When nucleation is difficult, corresponding to the interesting situations with long incubation times, the concentration of the metastable mother phase quickly reaches its maximum due to a balance of the incoming flux and desorption, and then slowly decreases due to nucleation (the so-called regime of incomplete condensation [22,44]). In this case, the concentration of material $k$ at the beginning of nucleation is given by $I_k^{des} c_k^\nu = I_k$. Here, $I_k^{des}$ is the desorption rate, $I_k$ is the incoming flux, with $\nu = 2$ for As desorbing in the form of As$_2$ molecules [45,46] and $\nu = 1$ for Ga atoms. The simplest approximation for the temperature dependence of the desorption rate has also the Arrhenius form $I_k^{des} = I_k^0 \exp(-E_k^{des}/k_B T)$, with $E_k^{des}$ as the activation energy for desorption, and a temperature-independent $I_k^0$ [44]. The inequality $c_k > c_k^{eq}$ is then reduced to

$$\ln(I_k / \widetilde{I}_k) + \frac{T_k}{T} > 0, \tag{1}$$

with the temperature-independent characteristic flux $\widetilde{I}_k = I_k^0 (c_k^0)^\nu$ and the characteristic temperature $T_k = (E_k^{des} + \nu \Lambda_k)/k_B$. This allows us to find the minimum flux and maximum temperature, both corresponding to the condition $c_k = c_k^{eq}$ when no nucleation occurs,

$$I_k^{\min} = \widetilde{I}_k \exp\left(-\frac{T_k}{T}\right), \quad T_k^{\max} = \frac{T_k}{\ln(\widetilde{I}_k / I_k)}. \tag{2}$$

Despite its simplicity, this treatment correctly reveals that the minimum flux increases with temperature, as in Fig. 1(a), and the maximum temperature increases with the flux.

A more detailed analysis of the GaAs nucleation relies upon the mononuclear regime for two-dimensional (2D) GaAs islands [47] of the first NW monolayer, yielding the nucleation time $t_{GaAs} = 1/(\pi R^2 J_{GaAs})$, where $R$ is the mean radius of the droplet base and $J_{GaAs}$ is the (Zeldovich) nucleation rate of 2D GaAs islands at the droplet-substrate interface [47,48]. According to Refs. [44,45,49,50], this nucleation rate can be presented in the form

$$J_{GaAs} = J_{GaAs}^0 \exp\left[-\left(\frac{E_{GaAs}^{surf}}{k_B T}\right)^2 \frac{1}{\ln(c_{As}/c_{As}^{eq})}\right], \tag{3}$$

in which $E_{GaAs}^{surf}$ is the appropriately normalized surface (or edge) energy of 2D GaAs islands.

For an ensemble of three-dimensional (3D) Ga droplets growing from the sea of Ga adatoms, we use the method of Ref. [51] at $\lambda_{Ga}^2 I_{Ga} t_{Ga}^{des} \ll 1$ in the regime of incomplete condensation, with $\lambda_{Ga}$ as the diffusion length of Ga adatoms and $t_{Ga}^{des}$ as their mean lifetime on the substrate surface before desorption [22,51]. On the other hand, we assume that $\lambda_{Ga}$ remains much larger than the mean radius of the droplet base $R$, in which case the droplet growth rate depends only logarithmically on $\lambda_{Ga}/R$ [44]. For the time-dependent nucleation rate of Ga droplets, we then have $J_{Ga}(t) = J_{Ga}^*/\cosh^2(t/t_{Ga})$ (Ref. [51]). Here, $J_{Ga}^*$ is the maximum nucleation rate corresponding to the maximum concentration of Ga adatoms, $t_{Ga} = \sqrt{c_{Ga}^{eq}\tau/(2\Gamma J_{Ga}^*)}$ is the nucleation time, $\tau = \ln(\lambda_{Ga}/R)/(2\pi D_{Ga} c_{Ga}^{eq})$ is the characteristic growth rate of Ga droplets, with $D_{Ga}$ as the diffusion coefficient of Ga adatoms on the surface [44]. These expressions predict that the total density of droplets $N_{Ga} = J_{Ga}^* t_{Ga}$ scales as $\sqrt{J_{Ga}^*}$ and hence is controlled by the same process as the nucleation of Ga droplets, as seen in Figs. 1 and 2. The maximum nucleation rate can be presented as [44,51]

$$J_{Ga}^* = J_{Ga}^0 \exp\left[-\left(\frac{E_{Ga}^{surf}}{k_B T}\right)^3 \frac{1}{\ln^2(c_{Ga}/c_{Ga}^{eq})}\right]. \qquad (4)$$

Here, $E_{Ga}^{surf}$ is the appropriately normalized surface energy of 3D Ga droplets resting on the substrate surface, $c_{Ga} = I_{Ga} t_{Ga}^{des}$ is the maximum concentration of the Ga adatoms, and $t_{Ga}^{des} = t_{Ga}^0 \exp(E_{Ga}^{des}/k_B T)$. Thus, the main differences between the incubation times for 2D GaAs islands emerging from 3D Ga droplets and Ga droplets themselves are in (i) different power indices under the exponents of Eqs. (3) and (4) due to 2D or 3D geometry and (ii) different dependences of the incubation times on the nucleation rates ($t_{GaAs} \propto 1/J_{GaAs}$ and $t_{Ga} \propto 1/\sqrt{J_{Ga}^*}$) in the mononuclear and polynuclear growth regimes, respectively.

Using the above expressions for $c_k$ and $c_k^{eq}$, it is easy to obtain the leading exponential temperature and flux dependences of the incubation times in the form

$$t_k = t_k^0 \exp\left[\frac{a_k}{\tau_k(1+\tau_k \ln i_k)^{\delta_k}}\right], \quad (5)$$

with $\tau_k = T/T_k$, $i_k = I_k/\tilde{I}_k$, $a_{GaAs} = 2[E_{GaAs}^{surf}/(E_{As}^{des} + 2\Lambda_{As})]^2$, $a_{Ga} = [E_{Ga}^{surf}/(E_{Ga}^{des} + \Lambda_{Ga})]^3/2$, $\delta_k = 1$ for GaAs and 2 for Ga. The total incubation time equals

$$t_{inc} = t_{Ga} + t_{GaAs}, \quad (6)$$

as discussed above. For the flux dependence at a given temperature, Eq. (5) is reduced to

$$t_k = t_k^0 \exp\left[\left(\frac{T_1^{(k)}}{T}\right)^{\delta_k+1} \frac{1}{\ln^{\delta_k}(I_k/I_k^{\min})}\right]. \quad (7)$$

Here, the parameters $T_1^{(GaAs)} = \sqrt{2}E_{GaAs}^{surf}/k_B$ and $T_1^{(Ga)} = E_{Ga}^{surf}/(2^{1/3}k_B)$ are related only to the surface energies of 2D GaAs islands and 3D Ga droplets, respectively. For the temperature dependence at a given flux, Eq. (5) is reduced to

$$t_k = t_k^0 \exp\left[\frac{T_0^{(k)}}{T(1-T/T_k^{\max})^{\delta_k}}\right], \quad (8)$$

with $T_0^{(k)} = a_k T_k$. Of course, this expression has physical meaning only for $T \leq T_k^{\max}$, with the discontinuity under the exponent at $T = T_k^{\max}$ corresponding to infinite incubation time. For any temperature above $T_k^{\max}$, there is no nucleation, meaning that the incubation time is infinite as well. We note that these incubation times correspond to the very beginning of the growth process, where the first objects (Ga droplets or GaAs monolayers under the droplets) nucleate on the substrate surface. The nucleation time of developed NWs, reflected in the width of their length distributions [33,48], may be much longer as discussed below.

The temperature dependence of the incubation time given by Eqs. (6) and (8) yields the excellent fits with the data shown by the solid lines in Fig. 1(b). The fitting parameters are $t_{Ga}^0 =$ 2.0 s, $T_{Ga}^{\max} = 699$ °C, and $T_0^{Ga} = 15.5$ K for the incubation time of Ga droplets, shown by the magenta line in the figure. $t_{Ga}^0$ and $T_0^{Ga}$ are constants characteristic for the derived dependencies but difficult to associate with exact physical meaning. The value of 699 °C for the maximum temperature at which nucleation is possible is consistent with the discussion of Fig. 1(b) in the previous section, but we point out that this value describes the divergence to infinity and should not be confused with the critical temperature introduced there. Using $j_{Ga} = 2\{E_{Ga}^{surf}/[k_B T \ln(c_{Ga}/c_{Ga}^{eq})]\}^3$ for the critical size and assuming that it equals 50 Ga atoms at 620 °C, we get $E_{Ga}^{surf} = 0.29$ eV. With this surface energy, we obtain the plausible estimate for $E_{Ga}^{des} + \Lambda_{Ga} = 3.1$ eV, consistent with Ref. [22]. It is seen that the incubation time for Ga droplets gives the main contribution into the total incubation time.

We have only one data point in Fig. 1(b) for the nucleation time of GaAs. However, combining it with the As flux dependence of the total incubation time at 620 °C shown in Fig. 1(a), the whole set of data can reasonably be fitted with $t_{As}^0 = 0.28$ s, $T_{GaAs}^{\max} = 690$ °C, $T_0^{GaAs} = 200$ K, $I_{As}^{\min} = 0.012$ atoms/nm$^2$/s and $E_{GaAs}^{surf} = 0.28$ eV. We note that the minimum As flux for which nucleation is possible as determined here is smaller than what we explored experimentally [cf. Fig. 1(a)]. However, $I_{As}^{\min}$ is the mathematical singularity, and incubation times may be impractically long already at higher As fluxes. The value for $E_{GaAs}^{surf}$ corresponds to the reasonable surface energy on the order of 1 J/m$^2$ (Ref. [49]). With this surface energy, the critical size $j_{GaAs} = \{E_{GaAs}^{surf}/[k_B T \ln(c_{As}/c_{As}^{eq})]\}^2$ (Ref. [44]) equals 14 GaAs pairs at $c_{As}/c_{As}^{eq} = 2.7$ and $T =$ 620 °C. This estimate seems plausible, because for well-developed NWs, i.e. nucleation below the droplet on the top facet of the GaAs NW, the critical size is usually smaller, only a few GaAs pairs [45,49,50], consistent with the fact that the initial NW nucleation on the substrate is difficult. We note, however, that the total incubation time decreases with increasing As flux to much shorter values than the incubation time for Ga droplets at $I_{As} = 2.8$ atoms/nm²/s ($\cong 22$ s). This suggests a decrease of the incubation time for Ga droplets toward higher $I_{As}$, which is accounted for by the simplest dependence $t_{Ga}^0 \propto 1/I_{As}$ in the theoretical curve in Fig. 1(a) at 620 °C. The curve at 580 °C in Fig. 1(a) is obtained with the same parameters except for $I_{As}^{min}$, which equals 0.0011 atoms/nm²/s in this case. This is consistent with the steep exponential temperature dependence given by Eq. (2). The incubation time for Ga droplets is simply put to zero at 580 °C.

Thus, our theoretical analysis not only quantitatively describes the observed behavior of the incubation times under different conditions, but also allows us to deduce otherwise unattainable physical parameters of the nucleation processes. Overall, we have shown that for Ga-assisted growth of GaAs NWs on Si substrates without Ga pre-deposition, the incubation time can be limited by either too high temperatures to nucleate Ga droplets or too low As fluxes to precipitate GaAs below these droplets. Furthermore, extremely steep flux and temperature dependences of the incubation times derived from the Zeldovich nucleation rate can approximately be fitted by the Arrhenius exponents only in limited domains. Outside these domains, the incubation times rapidly tend to infinity, showing that such growth conditions are inappropriate for reproducible growth of regular NW ensembles. There is no simple way to establish a quantitative correspondence between the detailed functions in Eqs. (6) and (7) and the activation energy in the approximate exponential fits, because the denominators under the exponents tend to infinity when $I_k$ approaches $I_k^{\min}$ or $T_k$ approaches $T_k^{\max}$. Physically transparent parameters of our process are the minimum arsenic flux and maximum temperature, for which nucleation of GaAs is suppressed due to excess desorption of either As or Ga species.

We emphasize that for all experiments presented here the same Ga flux was employed. In our previous study we observed that the total density of all objects depends on both T and Ga flux and can be kept fixed for an exponential relation between these two parameters [15]. This result suggests that the behaviors of the incubation time versus the As flux and temperature found in the present study should vary with the applied Ga flux, which is also supported by the model.

## V.  CONCLUSIONS

We have investigated the nucleation phenomena relevant for the Ga-assisted growth of GaAs NWs by MBE in different complementary ways, based on the *in-situ* measurement of the incubation time preceding NW formation. Our results show the importance of the appropriate choice of both As flux and growth temperature to avoid long nucleation delays and grow more regular ensembles of NWs. We have found a rather general effect of suppressing the NW nucleation at low As fluxes and high temperatures, and presented the threshold values for GaAs NWs. We believe that these results can be translated to other self-assisted III-V NWs and possibly even a broader range of surface nanostructures. In particular, the existence of a maximum temperature and minimum flux for obtaining regular ensembles of surface nanostructures must be a general phenomenon, and imposes important limitations on the choice of relevant growth conditions for a given material-substrate combination.

**ACKNOWLEDGMENTS**

The authors would like to thank A.-K. Bluhm for her SEM work and M. Höricke for MBE maintenance. The authors are grateful to V.M. Kaganer for a critical reading of the manuscript. This work was partially funded by Deutsche Forschungsgemeinschaft under grant Ge2224/2. VGD thanks the Russian Foundation for Basic Research for financial support under grants 17-52-16017, 18-02-40006, and 19-52-53031.